\PassOptionsToPackage{usenames}{xcolor}
\PassOptionsToPackage{dvipsnames}{xcolor}

\documentclass[sigconf]{acmart} 

\usepackage{subcaption}

\usepackage{enumitem}

\usepackage{booktabs} 

\usepackage{csquotes}
\usepackage{amsmath}
\usepackage{outlines}

\usepackage{eurosym}
\usepackage{siunitx}
\usepackage[usenames]{xcolor}

\usepackage{graphicx}

\usepackage{pifont}
\usepackage{array}
\newcolumntype{P}[1]{>{\centering\arraybackslash}m{#1}}
\newcolumntype{L}[1]{>{\arraybackslash}m{#1}}

\usepackage{xspace}
\newcommand\ie{i.\,e.\xspace}
\newcommand\eg{e.\,g.\xspace}

\newcommand\US{U.\,S.\xspace}

\newcommand{\mathup}[1]{\mathrm{#1}}

\usepackage{url}
\usepackage{balance}

\newcommand{\xmark}{\ding{55}}%

\makeatletter
\renewcommand{\fps@figure}{htb}         
\renewcommand{\fps@table}{htb}         
\makeatother

\allowdisplaybreaks

\usepackage{tabularx}

\AtBeginDocument{%
\providecommand\BibTeX{{%
\normalfont B\kern-0.5em{\scshape i\kern-0.25em b}\kern-0.8em\TeX}}}

\copyrightyear{2022}
\acmYear{2022}
\setcopyright{rightsretained}
\acmConference[WWW '22]{Proceedings of the ACM Web Conference 2022}{April 25--29, 2022}{Virtual Event, Lyon, France}
\acmBooktitle{Proceedings of the ACM Web Conference 2022 (WWW '22), April 25--29, 2022, Virtual Event, Lyon, France}
\acmDOI{10.1145/3485447.3512266}
\acmISBN{978-1-4503-9096-5/22/04}

\begin{document}

\fancyhead{}

\title{Moral Emotions Shape the Virality of COVID-19 \\Misinformation on Social Media}

\author{Kirill Solovev}
\email{kirill.solovev@wi.jlug.de}
\affiliation{
	\institution{JLU Giessen}
	\streetaddress{Licher Str.\ 62}
	\country{Germany}
}

\author{Nicolas Pröllochs}
\email{nicolas.proellochs@wi.jlug.de}
\affiliation{
	\institution{JLU Giessen}
	\streetaddress{Licher Str.\ 62}
	\country{Germany}
}

\renewcommand{\shortauthors}{Kirill Solovev and Nicolas Pröllochs}

\newcommand{\SUPPLEMENT}[1]{{Supplementary #1}}


\begin{abstract}
While false rumors pose a threat to the successful overcoming of the COVID-19 pandemic, an understanding of how rumors diffuse in online social networks is -- even for non-crisis situations -- still in its infancy. Here we analyze a large sample consisting of COVID-19 rumor cascades from Twitter that have been fact-checked by third-party organizations. The data comprises ${N =}$ 10,610 rumor cascades that have been retweeted more than 24 million times. We investigate whether COVID-19 misinformation spreads more viral than the truth and whether the differences in the diffusion of true vs.\ false rumors can be explained by the moral emotions they carry. We observe that, on average, COVID-19 misinformation is more likely to go viral than truthful information. However, the veracity effect is moderated by moral emotions: false rumors are more viral than the truth if the source tweets embed a high number of other-condemning emotion words, whereas a higher number of self-conscious emotion words is linked to a less viral spread. The effects are pronounced both for health misinformation and false political rumors. These findings offer insights into how true vs.\ false rumors spread and highlight the importance of considering emotions from the moral emotion families in social media content.
\end{abstract}

\ccsdesc[500]{Human-centered computing~Social media}
\ccsdesc[500]{Human-centered computing~Empirical studies in collaborative and social computing}
\ccsdesc[100]{Applied computing~Sociology}

\keywords{Social media, misinformation, COVID-19, virality, moral emotions, computational social science, explanatory modeling}

\maketitle

\section{Introduction}

Social media platforms play an ambivalent role during the COVID-19 pandemic. On the one hand, they represent an important source of health information for large parts of society \cite{Limaye.2020}. On the other hand, however, this crisis has bred a multitude of rumors \cite{Pennycook.2020b,Frenkel.2020,Gallotti.2020,Islam.2020,Kouzy.2020}, and verdicts of reputable fact-checking organizations (\eg, {politifact.com}, {snopes.com}) suggest that social media is rife with COVID-19 misinformation. COVID-19 misinformation on social media includes, but is not limited to, misinformation about vaccination, ``miracle cures,'' and supposed preventives \cite{Havey.2020}. False rumors can impact the timely and effective adoption of public health recommendations \cite{Waszak.2018}, the effectiveness of the countermeasures deployed by governments \cite{Rapp.2018}, and are sometimes even used as a political weapon \cite{Ricard.2020}. Given that exposure to misinformation frequently manifests in offline consequences \cite{Pennycook.2020b}, there is an urgency to study the spread of rumors on social media in the context of COVID-19. Tedros Adhanom Ghebreyesus, director-general of WHO, and other experts speak of an ``infodemic,'' which must be fought \cite{Zarocostas.2020}.

While previous research -- at least for non-crisis situations -- suggests that false rumors on social media tend to be more viral than the truth \cite{Vosoughi.2018,Prollochs.2021b}, the mechanism underlying its viral spread, though critical, remains unresolved. In this work, we approach this question through the lenses of morality and emotions and their role in rumor diffusion in polarized social media environments. Social media content delivers not only factual information but also carries moral ideas and sophisticated emotional signals \cite{Brady.2017}. Moral emotions provide the motivational force for humans to do good and to avoid doing bad \cite{Tangney.2007} and can even serve to ``moralize'' actions that would otherwise be considered non-moral \cite{Wheatley.2005}.
Since socially connected users often develop similar ideas and intuitions \cite{Limaye.2020,Brady.2017,Christakis.2009,Fowler.2008}, moral emotions are a key driver of information diffusion in polarized social media environments \cite{Brady.2017}. In the context of COVID-19, the overall discussion culture has repeatedly been characterized as highly polarized \cite{Cossard.2020,Jing.2021,Hart.2020,Allcott.2020,Druckman.2021,Havey.2020}. For instance, people have been observed to be divided in their perceptions of government responses, confidence in scientists, and support for protective actions \cite{Jing.2021,Hart.2020}. If COVID-19 rumors are highly polarizing to social media users, then the transmission of moral emotions likely plays a key role in the rumors' diffusion through social networks.


The principal moral emotions can be divided into two families \cite{Haidt.2003}. The families are the ``other-condemning'' family, comprising the emotions contempt, anger, and disgust, and the ``self-conscious'' family comprising the emotions shame, pride, and guilt \cite{Tracy.2004}. The former, other-condemning emotions, are reactions to the social behavior of others and involve a negative judgment or disapproval of others. In the context of morality, other-condemning emotions are sometimes also referred to as the ``hostility triad'' \cite{Rozin.1999}. Other-condemning emotions are typically associated with perceived moral violations, for example, in the context of individuals' rights and fairness \cite{VanStekelenburg.2017}. While the individual emotions in the other-condemning family (\ie, anger, contempt, and disgust) are often assumed to be not particularly explosive on their own, they can become a dangerous, explosive mix when compressed together \cite{Rozin.1999}. Their counterpart is the family of self-conscious emotions, which are evoked by self-reflection and self-evaluation. These emotions motivate individuals to behave in a socially acceptable fashion and are linked to prosocial behaviors such as empathy and altruism \cite{Rozin.1999,Haidt.2003}. As such, self-conscious emotions can enable social healing and avoid triggering the contempt, anger, and disgust of others \cite{Haidt.2003}. While previous research \cite{Brady.2017} broadly distinguished moral vs.\ non-moral emotions in social media content, we will investigate whether these two clusters of moral emotions (self-conscious vs.\ other-condemning emotions) have distinct effects on the diffusion of rumors in the context of COVID-19.


\textbf{Research hypothesis: }
In this work, we propose that the virality of true vs.\ false COVID-19 rumors can be explained by the moral emotions they carry. Although previous research suggests that false rumors are statistically more often retweeted \cite{Vosoughi.2018,Prollochs.2021b}, not every false rumor is necessarily more viral than a truthful rumor. Rather, misinformation going viral is oftentimes spread through echo chambers with exacerbated ideological polarization \cite{Choi.2020}. In these environments, ideological identity is more salient in guiding user behavior \cite{Weng.2013} and users are moved towards more extreme positions \cite{Cinelli.2021}. Polarization not only reduces verification behavior \cite{Kim.2019,Moravec.2019} but also makes users more receptive to hostility against others, \eg, for political attacks \cite{Tucker.2018}. Here other-condemning emotions embedded in the \emph{source tweets}, which start the rumor cascade, may function as accelerators and amplifiers \cite{VanStekelenburg.2017b}. In polarizing discussions about COVID-19, this would imply that radical ideas and beliefs are strengthened and are more likely to translate into action. Given increased ideological polarization for false rumors \cite{DelVicario.2019}, the explosive mix of other-condemning emotions should thus accelerate their spread within social networks. The same reasoning suggests that false rumors embedding self-conscious emotions (that avoid triggering other-condemning emotions \cite{Haidt.2003}) should be less contagious on social media. In sum, we hypothesize that rumors with a stronger combination of false content and other-condemning emotions in the source tweets reach more people, whereas the combination of false content and self-conscious emotions reaches fewer people. 

\textbf{Data: } We collected a \emph{unique} dataset of COVID-19 rumor cascades propagating on Twitter between January 2020 and the end of April 2021. Each rumor cascade was investigated and fact-checked by at least one of three independent fact-checking organizations (snopes.com, politifact.com, truthorfiction.com). Our data include \num{10610} rumor cascades that have been retweeted 24.34 million times.

\textbf{Methodology: } We use textual analysis to extract fine-grained moral emotions (self-conscious and other-condemning) embedded in rumor cascades. Specifically, we employ (and validated) a dictionary-based approach to count the frequency of occurrence of self-conscious and other-condemning emotion words in the source tweets that have initiated the rumor cascades. To measure the diffusion of each rumor cascade, we employ the Twitter Historical API to obtain the number of retweets, that is, the number of users interacting with the rumor cascade. We then fit \emph{explanatory} regression models to evaluate how variations in moral emotions are associated with differences in the number of retweets for true vs.\ false rumor cascades. In our regression analysis, we follow previous works \cite{Brady.2017,Vosoughi.2018} by controlling for variables known to affect the retweet rate independent of the main predictors, \ie, the number of followers, the account age, etc.

\textbf{Findings: } We observe that, on average, COVID-19 misinformation is more likely to go viral than truthful information. However, the veracity effect is moderated by moral emotions: false rumors are more viral than the truth if the source tweets embed a high number of other-condemning emotion words, whereas a higher number of self-conscious emotion words is linked to a less viral spread. The effects are pronounced both for health misinformation and false political rumors. These findings offer insights into how true vs.\ false rumors spread and highlight the importance of considering emotions from the moral emotion families in social media content.

\section{Background}

\subsection{Misinformation on Social Media}

Social media has shifted quality control for the content from trained journalists to regular users \cite{Kim.2019}. The lack of oversight from experts makes social media vulnerable to the spread of misinformation \cite{Shao.2016}. Social media has indeed repeatedly been observed to be a medium that disseminates vast amounts of misinformation \cite[\eg,][]{Vosoughi.2018,Prollochs.2022a}. The presence of misinformation on social media also has detrimental consequences on how opinions are formed in the offline world \cite{Allcott.2017,Bakshy.2015,DelVicario.2016,Oh.2013}. As a result, it not only threatens the reputation of individuals and organizations, but also society at large.


Several works have focused on the question of \emph{why} misinformation is widespread on social media. These studies suggest that it is difficult for users to spot misinformation as it is often intentionally written to mislead others \cite{Wu.2019}. Moreover, social media users are often in a hedonic mindset and avoid cognitive reasoning such as verification behavior \cite{Moravec.2019}. The vast majority of social media users do not fact-check articles they read \cite{Geeng.2020,Vo.2018}. A recent study further suggests that the current platform design may discourage users from reflecting on accuracy \cite{Pennycook.2021}. Online social networks are also characterized by (political) polarization \cite{Levy.2021,Prollochs.2022a,Solovev.2022a} and echo chambers \cite{Barbera.2015}. In these information environments with low content diversity and strong social reinforcement, users tend to selectively consume information that shares similar views or ideologies while disregarding contradictory arguments \cite{Ecker.2010}. These effects can even be exaggerated in the presence of repeated exposure: once misinformation has been absorbed, users are less likely to change their beliefs even when the misinformation is debunked \cite{Pennycook.2018}. 

\subsection{Research on Rumor Spreading}

Several studies have analyzed the spreading dynamics of rumors vs. non-rumors on social media. This includes analyses of summary statistics with regard to, for instance, the number of retweets \citep[\eg,][]{Bessi.2015,Friggeri.2014} and the rumor lifetime \cite[\eg,][]{Bessi.2015,Castillo.2011,DelVicario.2016}. However, these works discern cascades from rumors vs. non-rumors, and do not focus on differences across veracity. Another stream of literature has analyzed rumors concerning specific events (\eg, the 2013 Boston Marathon bombing) with regard to the overall tweet volume or content \citep[\eg,][]{Domenico.2013,Starbird.2014,Starbird.2017}. These works analyze how the user base responds to rumors but again do not analyze the diffusion dynamics of true vs. false rumors. 

Only a few works have analyzed differences in the spread of true vs. false rumors. Friggeri et al. (2014) \cite{Friggeri.2014} classified the veracity of $\approx$4,000 rumors from Facebook based on fact-checking assessments from snopes.com. The authors find that a majority of resharing of false rumors occurs after fact-checking. This suggests that social media users likely do not notice the fact-checks; or intentionally ignore their verdict. Closest to our work is the study from Vosoughi et al. (2018) \cite{Vosoughi.2018}, which provides a comprehensive analysis of summary statistics of true vs. false rumors on Twitter, finding that false rumors spread significantly farther, faster, and more broadly than the truth. However, this work does not analyze the spread of true vs. false rumors in the context of COVID-19. The same dataset \cite{Vosoughi.2018} has also been used in a recent study \citep{Prollochs.2021b} that measures emotions embedded in the \emph{replies} to rumor cascades. The authors find that higher frequencies of certain emotions (\eg, anger) are associated with more viral cascades for false rumors. 

In the context of COVID-19, research providing large-scale quantitative analyses of the spread of true and false rumors is scant. Existing works have primarily focused on summary statistics of small sets of hand-labeled rumors or source-based approaches to identify COVID-19 misinformation \cite[\eg,][]{Cinelli.2020,Kouzy.2020,Singh.2020}. For example, Cinelli et al. (2020) classify news sources into reliable and non-reliable sources in order to analyze the spread of COVID-19-related content. The authors find no significant differences regarding the spreading dynamics. Notably, however, 
categorizations of reliable vs. non-reliable sources do not necessarily correspond to true vs. false rumors. In addition, source-based approaches ignore false rumors from influential individuals, emerging websites, and misclassify false rumors from websites that are generally considered as being reliable.  
Note that there are other recent papers reporting that COVID-19 misinformation is widespread on social media, characterizing COVID-19 misinformation, and expressing concerns about consequences for public health \cite[\eg,][]{Griffith.2021,Gallotti.2020,Islam.2020,Kouzy.2020,Pennycook.2020b}. However, these works do not focus on modeling {differences in the diffusion} of true vs. false COVID-19 rumor cascades. 

\textbf{Our contributions: } This work makes two key contributions. (1) We collected a unique dataset of COVID-19 rumor cascades and demonstrate that misinformation is, on average, more viral than the truth. Here, our study connects to previous works \cite{Vosoughi.2018,Prollochs.2021b}, which yielded similar conclusions, yet not in the context of COVID-19. (2) The mechanisms underlying the viral spread of false rumors, though critical, have remained largely unresolved in previous research. Our work is the first to approach the question through the lenses of morality and emotions -- finding that moral emotions embedded in \emph{source tweets} shape the diffusion of false rumors on social media. 

\section{Methods}
\label{sec:methods}

\subsection{Data Collection}


\textbf{Fact-checks:} We identified three fact-checking organizations that thoroughly investigate rumors related to COVID-19. The names of the fact-checking organizations are: \url{politifact.com}, \url{truthorfiction.com}, and \url{snopes.com}. These fact-checking organizations list COVID-19 rumors in separate categories or tag them with a topic label (\eg, ``COVID-19'', ``Coronavirus'') which allows us to distinguish COVID-19-related rumors from other rumors (see Tbl.~\ref{tbl:fc_tags}). We scraped {all} COVID-19-related fact-checks from these platforms.

\begin{table}
\footnotesize
	\begin{center}
		\caption{Tags used to identify COVID-19-related fact-checks from fact-checking organizations.}
		\begin{tabular}{ll r}
			\toprule
			\textbf{Fact-checking organization} & \textbf{Tag} & \textbf{\#Fact-checks} \\
			\midrule
			politifact.org                      & Coronavirus   & 403                            \\
			snopes.com                          & COVID-19      & 265                            \\
			truthorfiction.com                  & covid-19      & 44                             \\ \bottomrule
		\end{tabular}
		\label{tbl:fc_tags}
	\end{center}
\end{table}

\begin{table*}
\footnotesize
	\setlength{\extrarowheight}{5pt}
	\begin{center}
		\caption{Summary statistics for tweets of rumor starters. Mean values are highlighted in bold, standard deviations are shown in parentheses. All Twitter variables were obtained from the Twitter Historical API.}
		\begin{tabularx}{\textwidth}{@{\hspace{\tabcolsep}\extracolsep{\fill}}l c c c c }
			\toprule
			{Variable}                & \multicolumn{1}{c}{{All cascades}}               & \multicolumn{1}{c}{\textsc{Politics}}            & \multicolumn{1}{c}{\textsc{Health}}              & \multicolumn{1}{c}{\textsc{Other}}               \\ \midrule
			Dates collected           & \textsf{01/02/20~--~05/13/21}                    & \textsf{01/02/20~--~05/13/21}                    & \textsf{01/27/20~--~05/13/21}                    & \textsf{01/27/20~--~05/12/21}                    \\
			Number of cascades        & \textsf{10,610}                                  & \textsf{8,157}                                   & \textsf{4,116}                                   & \textsf{1,297}                                   \\
			Number of retweets        & \textsf{24,339,625}                              & \textsf{20,374,097}                              & \textsf{10,231,382}                              & \textsf{1,416,474}                               \\
			Retweet count range       & \textsf{0~--~260,637}                            & \textsf{0~--~260,637}                            & \textsf{0~--~207,155}                            & \textsf{0~--~76,092}                             \\
			Proportion \emph{True}    & \textsf{35.3\%}                                  & \textsf{39.0\%}                                  & \textsf{34.7\%}                                  & \textsf{19.7\%}                                  \\
			Proportion \emph{False}   & \textsf{46.9\%}                                  & \textsf{42.7\%}                                  & \textsf{48.3\%}                                  & \textsf{64.8\%}                                  \\
			Proportion \emph{Mixed}   & \textsf{17.7\%}                                  & \textsf{18.3\%}                                  & \textsf{17.0\%}                                  & \textsf{15.6\%}                                  \\
			Followers                 & \textsf{\textbf{2,256,095} \textit{(7,700,566)}} & \textsf{\textbf{2,545,874} \textit{(8,260,175)}} & \textsf{\textbf{2,526,538} \textit{(9,166,450)}} & \textsf{\textbf{816,527.4} \textit{(3,859,619)}} \\
			Followees                 & \textsf{\textbf{9,193.9} \textit{(34,750.39)}}   & \textsf{\textbf{10,124.4} \textit{(37,507.33)}}  & \textsf{\textbf{9,320.23} \textit{(40,249.53)}}  & \textsf{\textbf{5,952.10} \textit{(25,541.03)}}  \\
			Account age               & \textsf{\textbf{3,333.35} \textit{(1,383.38)}}   & \textsf{\textbf{3,374.93} \textit{(1,376.82)}}   & \textsf{\textbf{3,321.95} \textit{(1,391.67)}}   & \textsf{\textbf{3,098.33} \textit{(1,386.78)}}   \\
			Verified users            & \textsf{55.1\%}                                  & \textsf{60.2\%}                                  & \textsf{56\%}                                    & \textsf{31.9\%}                                  \\
			Includes media            & \textsf{28.6\%}                                  & \textsf{27\%}                                    & \textsf{26.4\%}                                  & \textsf{38.2\%}                                  \\
			Other-condemning emotions & \textsf{\textbf{0.167} \textit{(0.217)}}         & \textsf{\textbf{0.164} \textit{(0.201)}}         & \textsf{\textbf{0.153} \textit{(0.190)}}         & \textsf{\textbf{0.189} \textit{(0.291)}}         \\
			Self-conscious emotions   & \textsf{\textbf{0.300} \textit{(0.209)}}         & \textsf{\textbf{0.294} \textit{(0.198)}}         & \textsf{\textbf{0.317} \textit{(0.196)}}         & \textsf{\textbf{0.321} \textit{(0.256)}}         \\ \bottomrule
		\end{tabularx}
		\label{tbl:descriptives}
	\end{center}
\end{table*}

The fact-checking organizations have different ways of labeling the veracity of a rumor. For example, politifact.com articles are given a ``Pants on Fire'' rating for false rumors, whereas snopes.com assigns a ``false'' label. Consistent with \citet{Vosoughi.2018}, we normalized the veracity labels across the different sites by mapping them to a score of 1 to 5. All rumors with a score of 1 or 2 were categorized as ``false,'' whereas rumors with a score of 4 or 5 were categorized as ``true.'' Rumors with a score of 3 were categorized as ``mixed.'' In some cases, the same rumors have been investigated by multiple fact-checking organizations. Previous research has shown that fact-checking websites show high pairwise agreement \citep{Vosoughi.2018}, ranging between \SI{95}{\percent} and \SI{98}{\percent}. Rumors classified as ``true'' or ``false'' even showed a perfect pairwise agreement of \SI{100}{\percent} \citep{Vosoughi.2018}. The resulting collection of fact-checks contained the following information: (i) the veracity label (``true'', ``false'', ``mixed''), (ii) links to the articles of the fact-checking organizations, and (iii) the headline of the article that is being verified. 

\textbf{Rumor cascades on Twitter: }
We followed the approach from Vosoughi et al. (\citeyear{Vosoughi.2018}) to identify rumor cascades on Twitter: A rumor cascade on Twitter starts with a user making an assertion about a topic such as tweeting a text message or a link to an article. Social media users then propagate the rumor by retweeting it. Oftentimes, people also reply to the original tweet. These replies sometimes contain links to fact-checking organizations that either confirm or debunk the rumor in the original tweet. We used such cascades to identify rumor cascades that are propagating on Twitter.

We employed the Twitter Historical API to map the rumors to retweet cascades on Twitter as follows. First, we collected all tweets that contain a link to any of the websites from the fact-checking organizations. Second, for each reply tweet, we extracted the original tweet and the number of retweets of the original tweet. Here, special care is needed to ensure that the replies containing a link to any of the trusted websites address the original tweet. We followed the approach from \citet{Vosoughi.2018} to address this important issue: (i) we considered only replies to the original tweet and exclude replies to replies. (ii) To ensure that we study how unverified and contested information diffuses on Twitter, we removed all original tweets that are directly linking to one of the fact-checking websites. Note that tweets linking to one of the fact-checking websites do not qualify as they are no longer unverified. (iii) We compared the headline of the linked article to that of the original tweet. For this purpose, we used Universal Sentence Encoder \citep{Cer.2018} to convert the headline of the fact-check and the original tweet to vector representations that capture their semantic content. We then used cosine similarity to measure the distance between the vectors. If the cosine similarity was lower than 0.4, the tweet was discarded. 

The retweet cascades remaining after these filtering steps then represent rumors propagating on Twitter -- for which a veracity label is known based on the assessment from the fact-checking organization. 
In our data, the frequencies of fact-checking labels at cascade level are: \num[group-separator={,},group-minimum-digits=1]{3748} ($=$true), \num[group-separator={,},group-minimum-digits=1]{4979} ($=$false), and \num[group-separator={,},group-minimum-digits=1]{1883} ($=$mixed). These \num[group-separator={,},group-minimum-digits=1]{10610} rumor cascades have received more than 24.33 million retweets by Twitter users. 

Following previous works \cite{Brady.2017,Vosoughi.2018}, we employed the Twitter API to collect a set of additional user variables for each source tweet, \ie, the number of followers, the account age, etc. These variables are known to affect the retweet rate and are later used as control variables in our regression model. Summary statistics of our dataset are reported in Tbl.~\ref{tbl:descriptives}.

\subsection{Calculation of Emotion Scores}

The ``other-condemning'' family of moral emotions comprises the emotions \emph{anger}, \emph{disgust}, and \emph{contempt}, whereas the ``self-conscious'' family comprises the emotions \emph{shame}, \emph{pride}, and \emph{guilt} \citep{Tracy.2004,Haidt.2003}. We employed text mining methods to measure the extent to which these emotions are embedded in the source tweets. For this purpose, we first applied standard preprocessing steps from text mining. Specifically, the running text was converted into lower-case and tokenized, and special characters (\eg, hashtags, emoticons) were removed. Subsequently, we applied (and validated) a dictionary-based approach analogous to earlier research \citep{Brady.2017, Vosoughi.2018,Prollochs.2021b}.

We measured other-condemning and self-conscious emotions embedded in the source tweets based on the NRC emotion lexicon \citep{Mohammad.2013}. This lexicon comprises 181,820 English words that are classified according to the emotions of Plutchik's emotion model \citep{Plutchik.1984}. Plutchik's emotion model defines 8 basic emotions and 24 emotional dyads. The emotional dyads represent complex emotions, which are derived as a combination of two basic emotions \cite{Prollochs.2021a}. 
We used the NRC dictionary to count the frequency of words in the tweets that belong to each of the emotions. Afterwards, we divided the word counts by the total number of dictionary words in the text, so that the vector is normalized to sum to one across the emotions \citep{Vosoughi.2018,Prollochs.2021b}. In our data, 78.15\% of all source tweets contained at least one emotion word from the NRC lexicon.
We filtered out tweets that do not contain any emotional words since, otherwise, the denominator is not defined \citep{Vosoughi.2018,Prollochs.2021b}. However, our later analysis yields qualitatively identical results when including these observations (\ie, assigning zero values). Based on the scores for the 8 basic emotions and the 24 derived emotions, and the definitions of the two moral emotion families \citep{Tracy.2004}, we calculated other-condemning emotions by taking the sum of \emph{anger}, \emph{disgust}, and \emph{contempt}. Self-conscious emotions were calculated by taking the sum of \emph{shame}, \emph{pride}, and \emph{guilt}. 

\textbf{User study: } In order to test the construct validity of our dictionary-based approach, we employed two trained research assistants to annotate a random subset of 200 tweets that were categorized as being more other-condemning than self-conscious based on the dictionaries; and a random subset of 200 tweets that were categorized as being more self-conscious than other-condemning. For each of the 400 tweets, the annotators were asked to what extent the tweet relates to other-condemning and self-conscious emotions on two 5-point Likert scales, ranging from 1 (``not related to [other-condemning, self-conscious] emotions at all'') to 5 (``very related to [other-condemning, self-conscious] emotions''). The annotators viewed the tweets in randomized order and were explained the difference between other-condemning and self-conscious emotions. {The annotators exhibited a statistically significant inter-rater agreement according to Kendall's $W$ ($p < 0.01$).} Furthermore, the annotators rated the random subset of other-condemning tweets as more ``other-condemning'' than ``self-conscious'' {$[t = 6.53, p < 0.001]$}; and the random subset of self-conscious tweets as more ``self-conscious'' than ``other-condemning'' {$[t = 4.50, p < 0.001]$}.

\begin{table*}
	\centering
	\footnotesize
	\renewcommand{\arraystretch}{1.7}
	\caption{Exemplary tweets of rumor starters for each topic.}\label{tbl:example_tweets}
	\begin{tabular}{llp{15cm}}
		\toprule
		{Topic}           & {Veracity} & {Twitter Message}                                                                                                                                                     \\
		\midrule
		\textsc{Politics} & {True}     & Trump fired the Pandemic response team in 2018... He did not replace them... \#TrumpYoureKilling \\
		\textsc{Politics} & {False}    & Sick: Nancy Pelosi tried to insert abortion funding measures into the Chinese Coronavirus response stimulus package I never want to hear that Donald Trump is politicizing this pandemic again while Democrats try this stunt This is a disgrace—Speaker Pelosi should be ashamed \\
		\textsc{Health}   & {True}     & More police officers have died from Covid-19 this year than have been killed on patrol. Gunfire is the second-highest cause of death. \\
		\textsc{Health}   & {False}    & 80\% of People Taking Maderna Vaccine Had Significant Side-Effects. While the killer Bill Gates laughs all the way to the bank. Stop this insanity now! \\
		\textsc{Other}    & {True}     & This is the first day of school in Paulding County, Georgia.\\
		\textsc{Other}    & {False}    & I thought this was supposed to be a conspiracy theory. But here it is, straight from Trudeau’s mouth. The pandemic is the excuse for a “Great Reset” of the world, led by the UN.\\
		\bottomrule
		\label{tbl:emotion_examples}
	\end{tabular}
\end{table*}

\subsection{Rumor Topics}

We employed a weakly supervised machine learning framework \citep{Yao.2020} to infer the topics in the source tweets that have initiated the rumor cascades. The benefit of this state-of-the-art approach is that (i) it is regarded as superior to conventional topic modeling (\ie, Latent Dirichlet Allocation) for short texts \citep{Yao.2020}, and (ii) its weakly supervised nature allows for an ex-ante selection of topics that we perceive as being particularly relevant in the context of COVID-19. 
We categorized the rumor cascades into three (not mutually exclusive) topics: \textsc{Health} (\eg, rumors about the safety of vaccines), \textsc{Politics} (\eg, allegations of political opponents), and \textsc{Other} (\ie, rumors that do not fall into one of the other categories). Example tweets for each topic are provided in Tbl.~\ref{tbl:example_tweets}.

Our weakly supervised machine learning framework proceeded in three steps (see \citet{Yao.2020} for methodological details): (1)~We started to identify topic-related tweets based on a set of manually selected keywords for each topic. For instance, for the topic \textsc{Health}, we searched for all tweets containing words such as \enquote{vaccine,} \enquote{flu,} \enquote{mask,} etc. (see list of keywords in the Supplementary Materials). (2)~We conducted clustering-assisted manual word sense disambiguation on the keyword-identified tweets \citep{Yao.2020}. Here we used the $k$-means clustering algorithm with Silhouette criterion to cluster the keyword-identified tweets for each topic. We then manually inspect random tweets sampled from each cluster and assessed whether the tweets in the cluster refer to the topic. We excluded each tweet cluster that does not show the pertinent meaning of the topic keyword. This allowed us to significantly clean and improve the quality of the keyword-identified tweets. (3)~We used the created labeled data to train a deep neural network classifier and learn to predict whether or not individual Twitter messages belong to a certain topic. The input data for the training machine learning classifier was a vector representation of the (cleaned) keyword-identified tweets and the topic label. To create vector representations of tweets, we used neural language models in the form of the Universal Sentence Encoder \citep{Cer.2018}. In our deep neural network classifier, we treated the task of predicting topic labels for (vector representations of) tweets as a multi-label problem considering that one tweet may belong to multiple topics (\ie, \textsc{Health} and \textsc{Politics}). In training, we used an equal number of \num{1000} keyword-identified Tweets for each topic as positive training instances. In addition, we used the excluded tweets from step (2) and randomly sampled unlabeled tweets equal to the sum of labeled tweets as negative training instances, \ie, with a topic label \textsc{Other}.

\textbf{User study: }
To ensure that the topic predictions are accurate, we tested for the presence of errant tweets with the help of two trained research assistants. We randomly sampled 200 tweets for each topic, and instructed the research assistants to annotate the tweets. Each annotator was asked to judge the validity of the topic label on a 5-point Likert scale, ranging from 1 (``not related to [topic] at all'') to 5 (``very related to [topic]''). When comparing the human annotations to the predicted topic labels, we found very few misclassified instances. On average, the share of tweets that were not classified as at least ``somewhat related to  [topic]'' was lower than {\SI{8.5}{\percent}} (see Supplementary Materials). 

\begin{figure*}[ht!]
	\captionsetup{position=top}
	\centering
	\subfloat[]{\includegraphics[width=.30\linewidth]{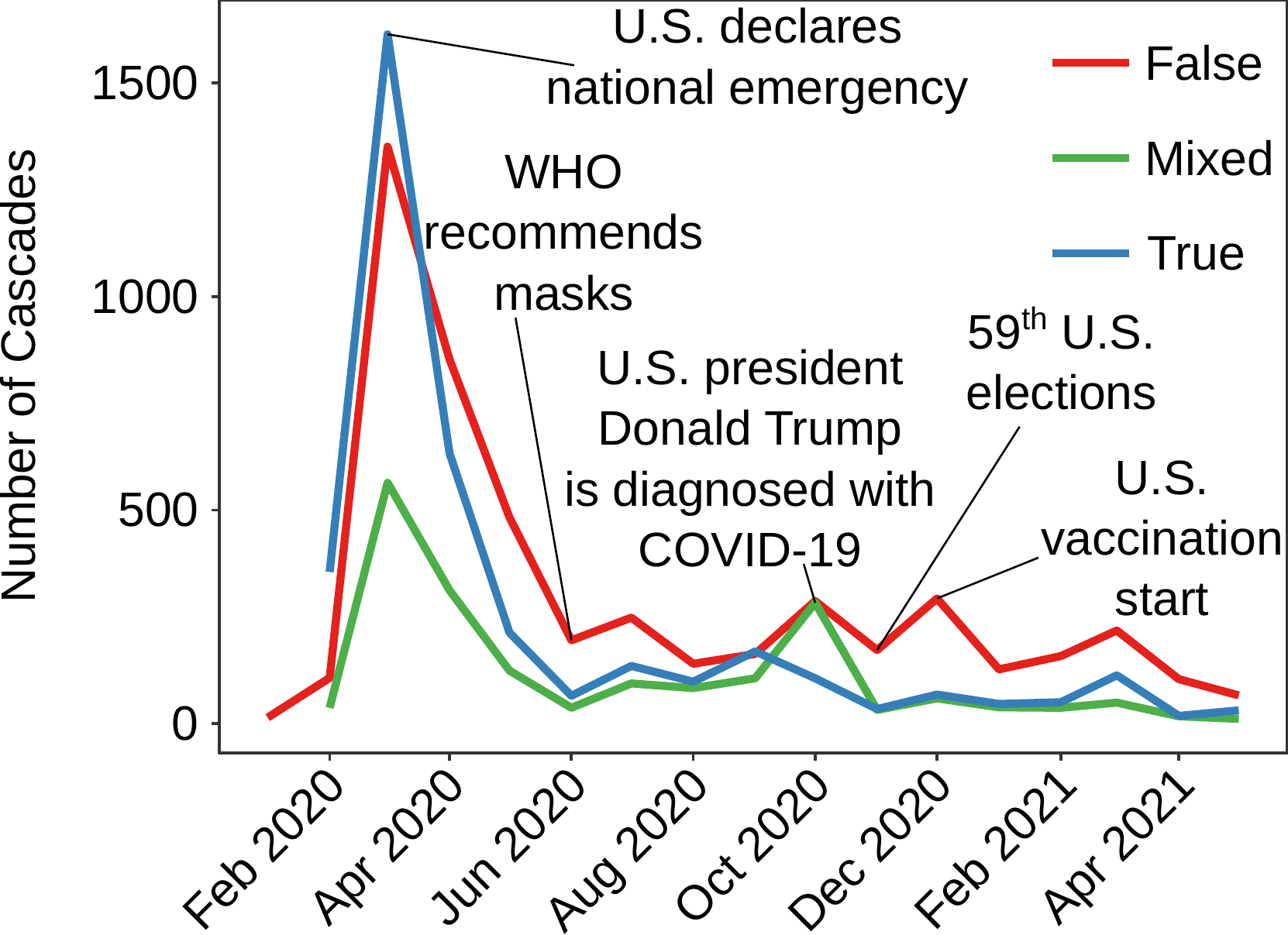}\label{fig:Annotated_monthly_cascades}}
	\hspace{0.4cm}
	\subfloat[]{\includegraphics[width=.30\linewidth]{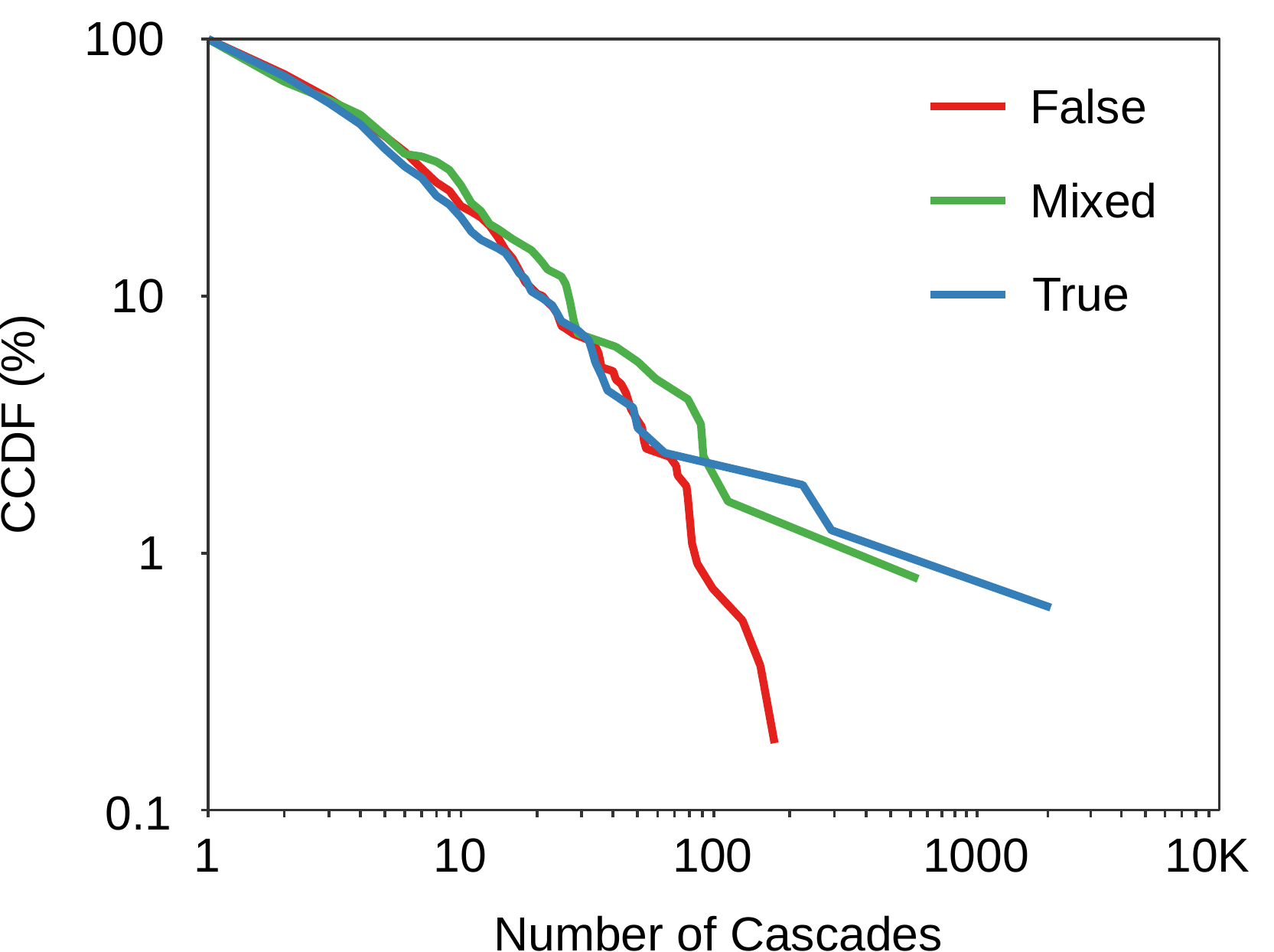}\label{fig:ccdf_cascade_number}}
	\hspace{0.4cm}
	\subfloat[]{\includegraphics[width=.30\linewidth]{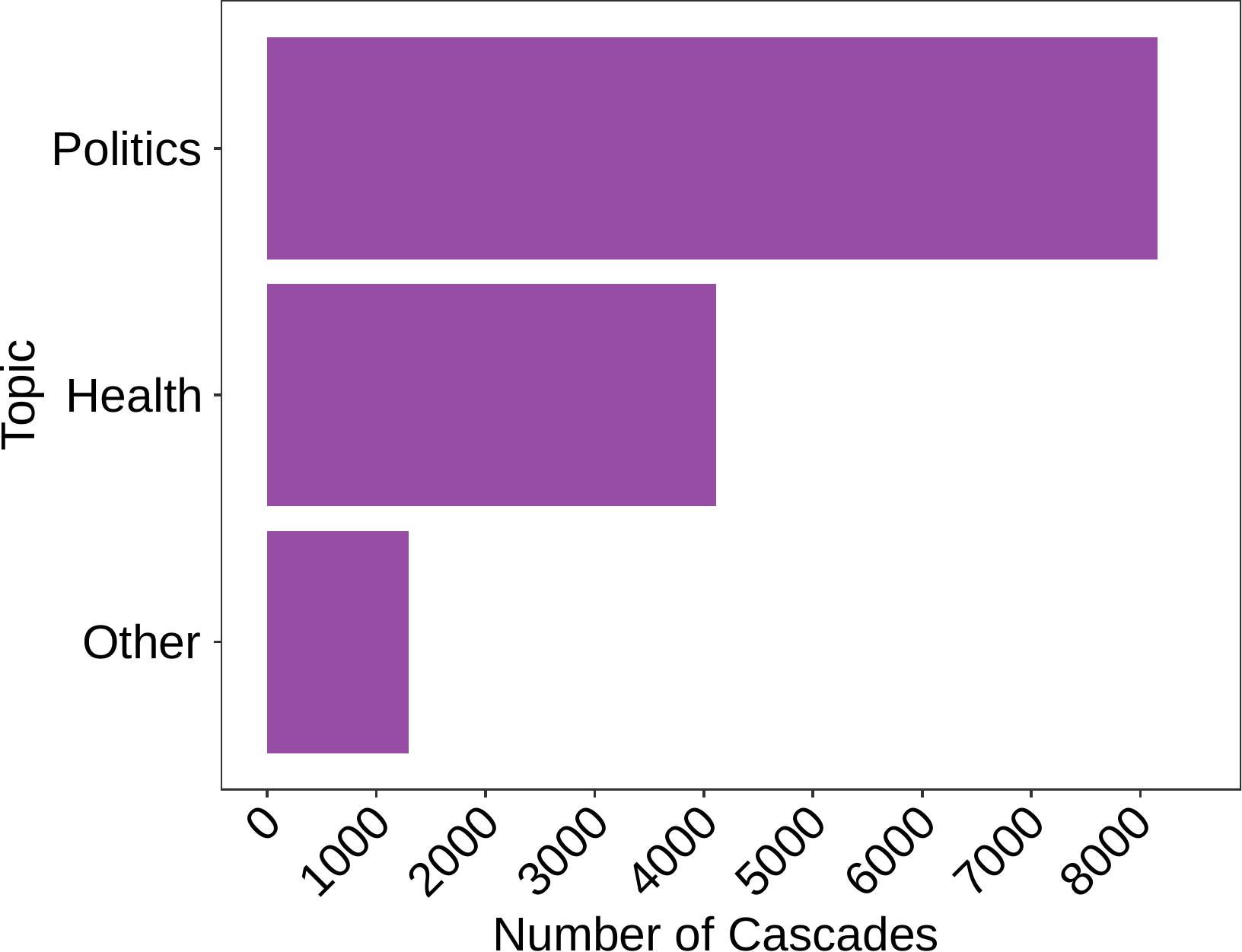}\label{fig:bar_topics}}
	\vspace{.25cm}
	\caption{COVID-19 rumor cascades propagating on Twitter between January 2020 and the end of April 2021. (A) Monthly counts of true, false, and mixed rumor cascades. (B) Complementary cumulative distribution functions (CCDFs) of true, false, and mixed rumor cascades. (C) {Number of rumor cascades across different topics.}}
	\label{fig:cascades}
\end{figure*}

\subsection{Model Specification}

We specified regression models with interaction terms that explain the number of retweets based on rumor veracity and other-condemning emotions and self-conscious emotions. Let $\mathit{RetweetCount_{i}}$ denote the number of retweets for rumor cascade $i$. Furthermore, let $\mathit{OtherCondemning_i}$ denote the proportion of other-condemning emotions, $\mathit{SelfConscious_i}$ the proportion of self-conscious emotions, and $\mathit{Falsehood_i}$ the veracity. Here we define a true rumor as $Falsehood_i = 0$ and a false rumor as $\mathit{Falsehood_i = 1}$. We adjusted for variables known to affect retweet rate \cite{Stieglitz.2013,Brady.2017,Vosoughi.2018,Prollochs.2021a,Prollochs.2021b,Prollochs.2022a}, which included the number of followers ($\mathit{Followers_i}$) and followees ($\mathit{Followees_i}$) of the author of the tweet, the account age ($\mathit{AccountAge_i}$), whether the author was verified by Twitter ($\mathit{Verified_i}$), and whether media was attached to the tweet ($\mathit{HasMedia_i}$). Each of these factors was extracted from the Twitter API. We $z-$standardized all continuous predictors in order to facilitate interpretability.

Based on the above variables, we specified the following generalized linear model for our analysis:
\begin{align}
     & \log({\mathup{E}(RetweetCount_i \,\mid\, ^*)}) = \, \beta_0 + \beta_{1} \, \mathit{Falsehood_i}  \nonumber                   \\
     & \qquad + \beta_{2} \, \mathit{Falsehood_i} \times \mathit{OtherCondemning_i}  \nonumber                                      \\
     & \qquad + \beta_{3} \, \mathit{Falsehood_i} \times \mathit{SelfConscious_i}  \nonumber                                        \\
     & \qquad + \beta_{4} \mathit{OtherCondemning_i} + \beta_{5} \, \mathit{SelfConscious_i}
    \label{eq:regression_volume}                                                                                                \\
     & \qquad + \beta_{6} \, \mathit{Followers_i} + \beta_{7} \, \mathit{Followees_i} + \beta_{8} \, \mathit{Account Age_i} \nonumber \\
     & \qquad + \beta_{9} \, \mathit{HasMedia_i} +  \beta_{10} \, \mathit{Verified_i}                                               
    \nonumber
\end{align}
with intercept $\beta_0$. 


$\mathit{RetweetCount}$ is a non-negative count variable, and its variance is larger than the mean. To adjust for overdispersion, we drew upon a negative binomial regression \cite{Stieglitz.2013,Brady.2017}. Note that because we estimate a negative binomial regression model with interaction terms, the coefficients cannot be interpreted as the change in the mean of the dependent variable for a one unit (\ie, standard deviation) increase in the respective predictor variable, with all other predictors remaining constant.
The reason is that in nonlinear regression models with interaction terms, marginal effects are nonlinear functions of the coefficients and the levels of the explanatory variables \citep{Buis.2010}. Instead, the coefficients can be interpreted on a multiplicative scale by calculating the incidence rate ratio (IRR), which is equal to the exponent of the coefficient of the respective variable \citep{Buis.2010}. Here the coefficients can be interpreted as the natural logarithm of a multiplying factor by which the predicted number of retweets changes, given a one unit increase in the predictor variable, holding all other predictor variables constant \citep{Buis.2010}.

\section{Results}


Our data include \num{10610} rumor cascades that have been retweeted 24.34 million times. The total number of COVID-19 rumor cascades peaked in March 2020 when the \US government declared a national emergency concerning the coronavirus disease and again in October 2020, the month prior to the \US presidential elections (Fig.~\ref{fig:Annotated_monthly_cascades}). The three fact-checking organizations have categorized \SI{46.9}{\percent} of all rumors as false, \SI{35.3}{\percent} as true, and \SI{17.7}{\percent} as being of mixed veracity. While the absolute number of rumor cascades has decreased over the course of the pandemic, the relative share of false vs.\ true rumors has increased (Fig.~\ref{fig:Annotated_monthly_cascades}). Compared to false rumors, a greater fraction of true rumors experienced more than 100 rumor cascades (Fig.~\ref{fig:ccdf_cascade_number}).
COVID-19 rumors are not constrained exclusively to health topics (\eg, rumors about the safety of vaccines). Rather, a sizable number of COVID-19 rumors concern political topics (\eg, true or false allegations of political opponents) \cite{Cossard.2020}. We thus applied topic modeling to categorize the rumor cascades in our dataset into three (not mutually exclusive) topics: \textsc{Politics}, \textsc{Health}, and \textsc{Other}. Fig.~\ref{fig:bar_topics} shows that a large proportion of COVID-19 rumors were thematically related to \textsc{Politics} (\SI{76.9}{\percent}), \textsc{Health} (\SI{38.8}{\percent}), while only \SI{12.2}{\percent} concerned \textsc{Other} topics (\eg, conspiracy theories). A total share of \SI{34.1}{\percent} of rumor cascades were thematically related to both \textsc{Politics} and \textsc{Health}.

\begin{figure*}
	\captionsetup{position=top}
	\centering
	\subfloat[]{\includegraphics[width=.30\linewidth]{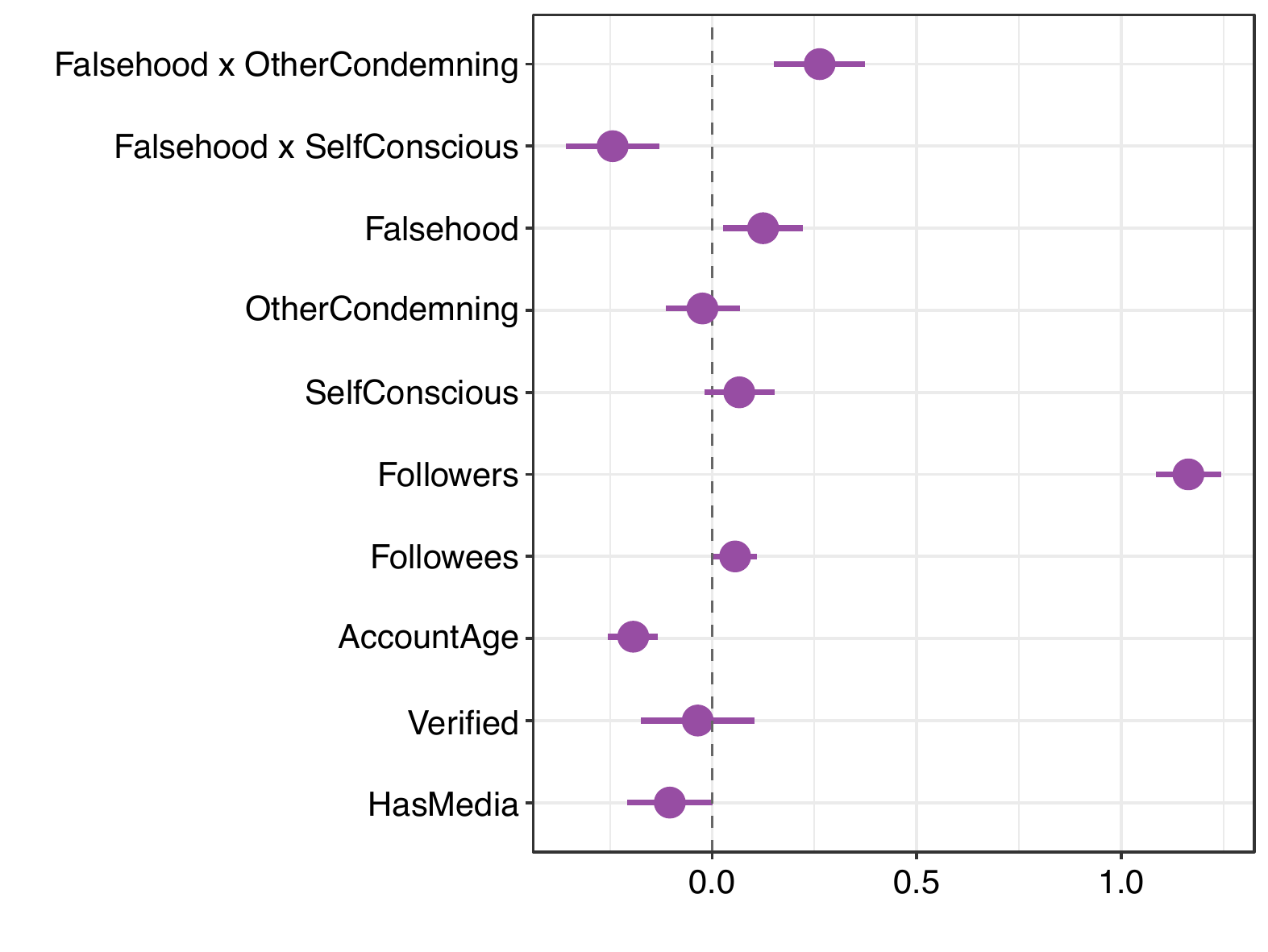}\label{fig:coefs_moral_full}}
	\hspace{0.4cm}
	\subfloat[]{\includegraphics[width=.30\linewidth]{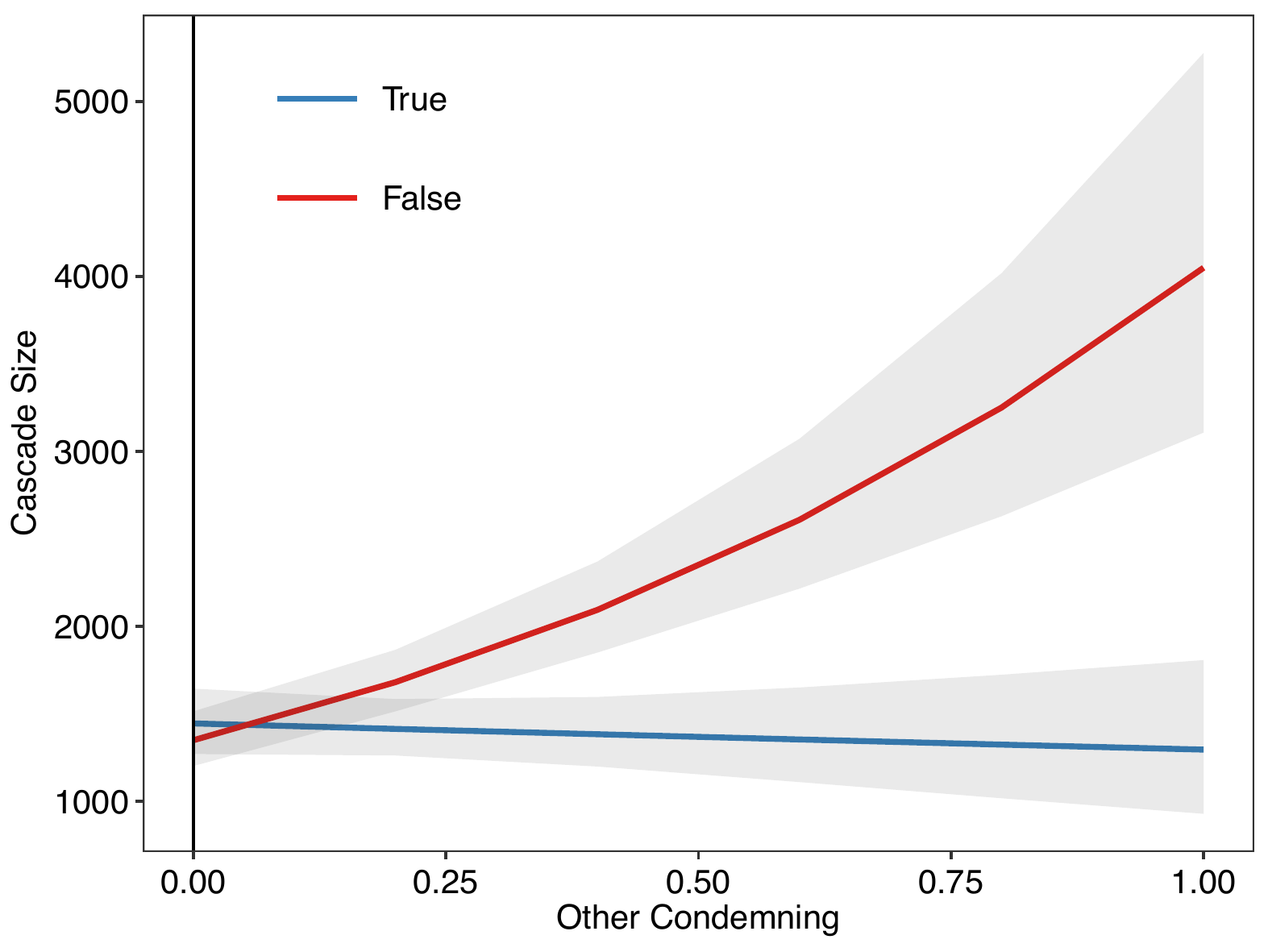}\label{fig:ggpredict_other_condemning}}
	\hspace{0.4cm}
	\subfloat[]{\includegraphics[width=.30\linewidth]{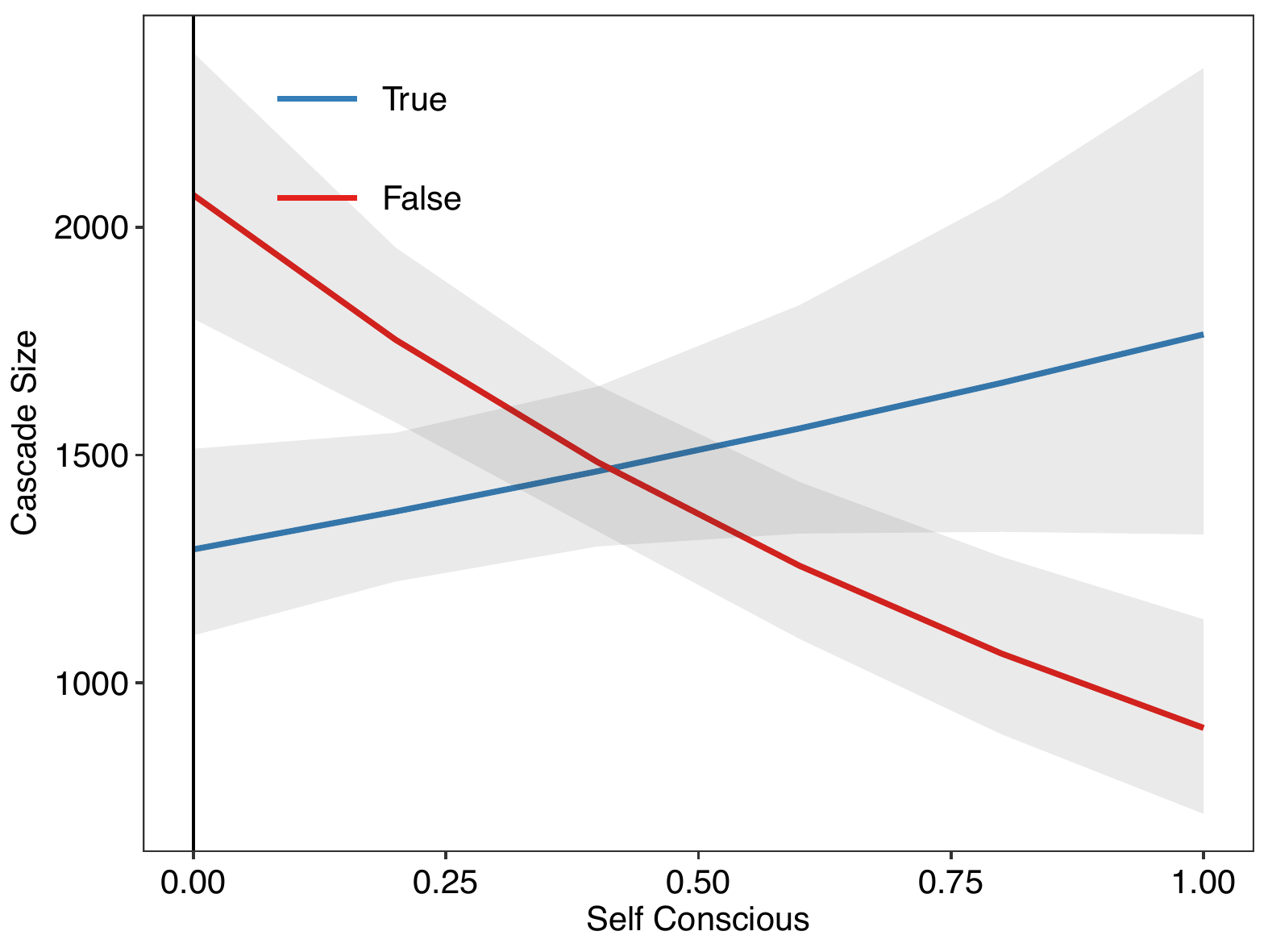}\label{fig:ggpredict_self_conscious}}
	\vspace{.25cm}
	\caption{Increases in other-condemning emotions predict higher retweet counts for false rumors, whereas increases in self-conscious emotions predict less retweets. (A) Coefficient estimates for negative binomial regression with \SI{95}{\percent} confidence intervals. The dependent variable is the number of retweets. (B--C) Predicted marginal means of the number of retweets for other-condemning emotions and self-conscious emotions. The \SI{95}{\percent} confidence intervals are highlighted in gray.
	}
	\label{fig:marginal_means_moral_emotions}
\end{figure*}


\textbf{Regression analysis: }
We fitted explanatory regression models to evaluate how variations in moral emotions are associated with differences in the number of retweets for true vs.\ false rumor cascades.
In our regression analysis, we followed previous works \cite{Brady.2017,Vosoughi.2018} by controlling for variables known to affect the retweet rate independent of the main predictors, \ie, the number of followers, the account age, etc.

As a baseline, we started our regression analysis with a negative binomial regression explaining the number of retweets solely based on the veracity label and control variables (see Supplementary Materials). Here false rumors (Falsehood $=1$) were estimated to receive \SI[retain-explicit-plus]{15.66}{\percent} more retweets than true rumors (IRR 1.16; $p<0.01$). 
Subsequently, we extended the negative binomial regression by including interaction terms between rumor veracity and other-condemning emotions, and between rumor veracity and self-conscious emotions (Fig.~\ref{fig:coefs_moral_full}). The coefficient estimates for these two interaction terms were statistically significant, which implies that false rumors' virality depended on the moral emotions embedded in the source tweet. Specifically, a one standard deviation increase in other-condemning emotions for false rumors was linked to a \SI[retain-explicit-plus]{26.99}{\percent} increase in the number of retweets (IRR 1.27; $p<0.01$). In contrast, a one standard deviation increase in self-conscious emotions for false rumors was linked to a \SI[retain-explicit-plus]{23.43}{\percent} decrease in the number of retweets (IRR 1.23; $p<0.01$). We found no statistically significant effect of other-condemning and self-conscious emotion words for true rumors. In sum, we observed that false rumors were more viral than the truth if the source tweet embedded a high proportion of other-condemning emotion words, whereas a high proportion of self-conscious emotion words was linked to a less viral spread (see Fig.~\ref{fig:ggpredict_other_condemning},~\ref{fig:ggpredict_self_conscious}).

\begin{figure*}
	\captionsetup{position=top}
	\centering
	\subfloat[]{\includegraphics[width=.30\linewidth]{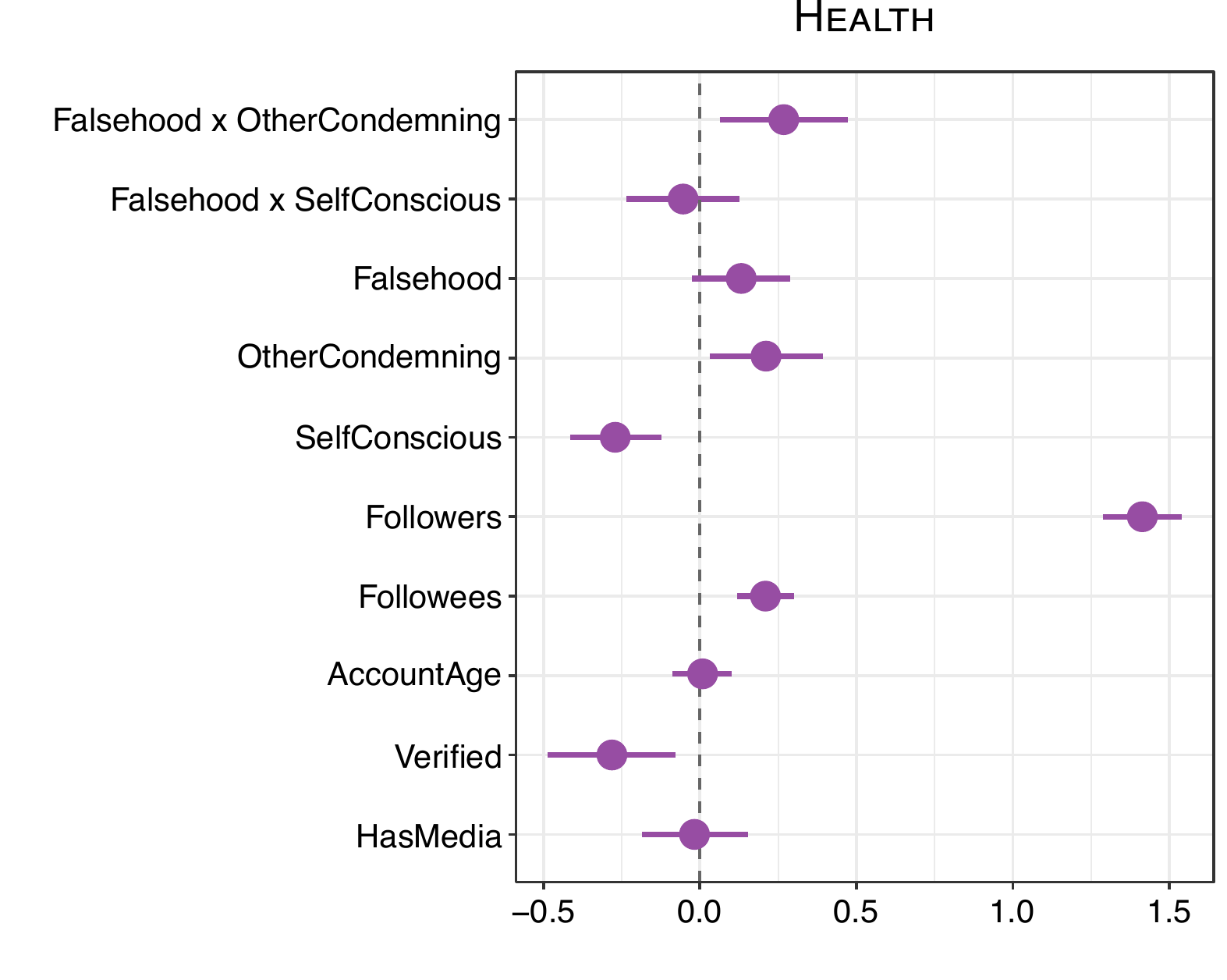}\label{fig:coefs_moral_health}}
	\hspace{0.4cm}
	\subfloat[]{\includegraphics[width=.30\linewidth]{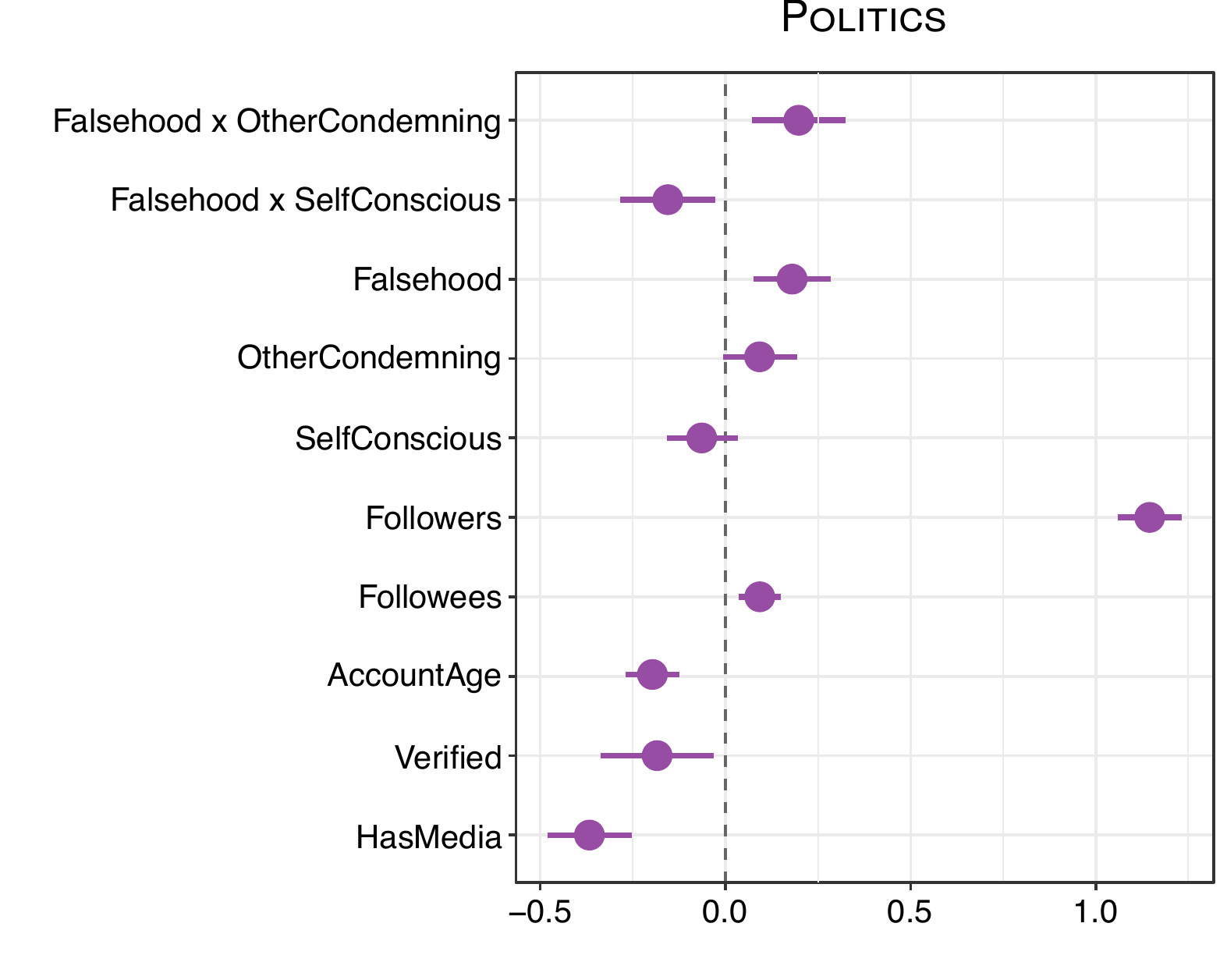}\label{fig:coefs_moral_politics}}
	\hspace{0.4cm}
	\subfloat[]{\includegraphics[width=.30\linewidth]{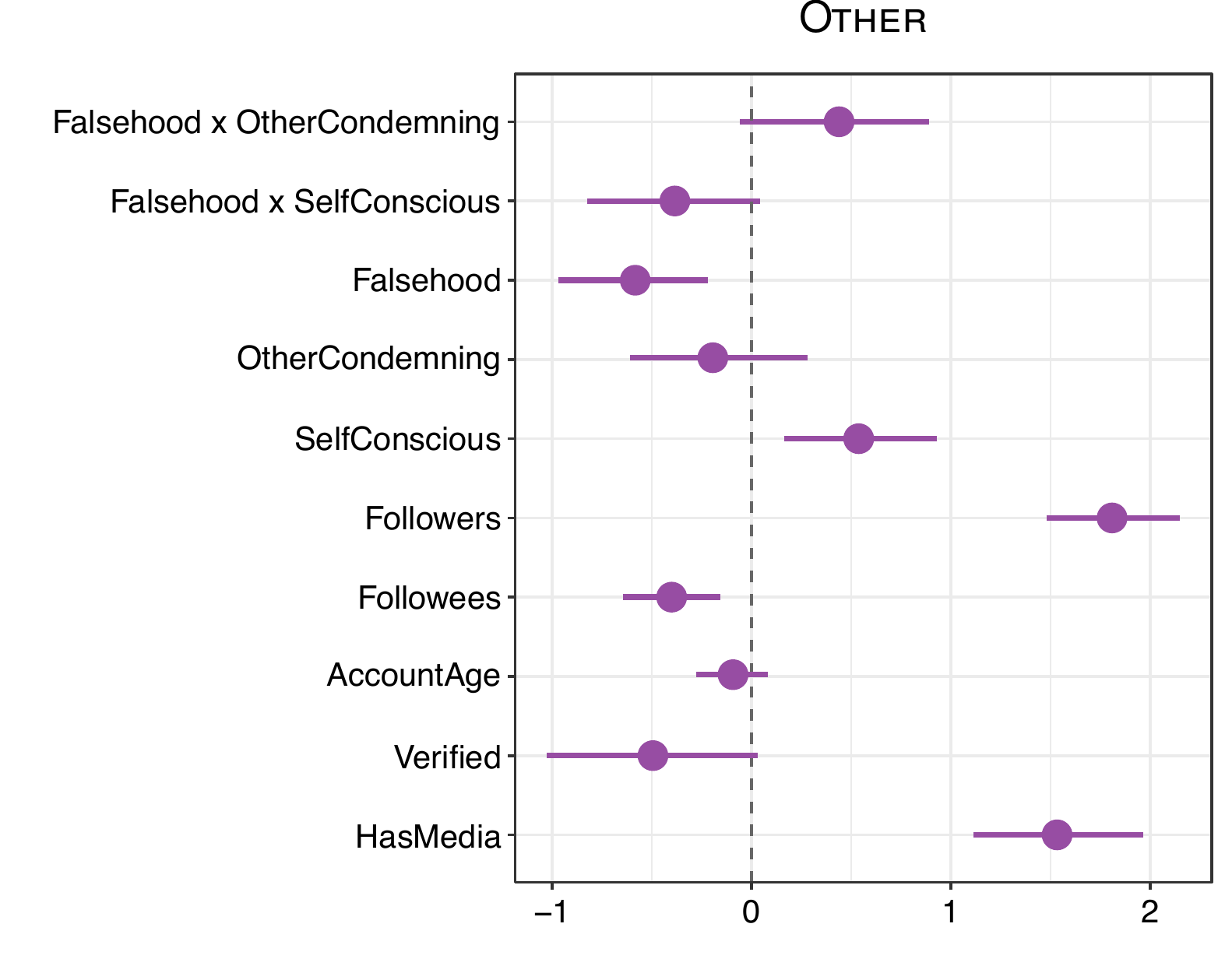}\label{fig:coefs_moral_origins}}
	\vspace{.15cm}
	\caption{Coefficient estimates for negative binomial regressions with \SI{95}{\percent} confidence intervals for rumor cascades filtered by topic (\emph{A:} \textsc{Health}, \emph{B:} \textsc{Politics}, and \emph{C:} \textsc{Other}). The dependent variable is the number of retweets.
	}
	\label{fig:marginal_means_topics}
\end{figure*}

\textbf{Analysis across topics: }
We also examined the effect of moral emotions across different topics. For each topic from Fig.~\ref{fig:bar_topics}, we generated observation subsets and re-estimated our regression model (Fig.~\ref{fig:marginal_means_topics}). We observed differences in the effects of moral emotions on the number of retweets. The effect of other-condemning emotions on the number of retweets was pronounced both for false rumors from the \textsc{Health} category (IRR 1.34; $p<0.01$) and for false rumors from the \textsc{Politics} category (IRR 1.61; $p<0.01$). For the \textsc{Other} category (with comparatively smaller sample size), the coefficients pointed in the same directions but were not statistically significant at common significance thresholds.


\textbf{Analysis of deleted tweets: } 
Major social media platforms such as Twitter have intensified their efforts to combat the spread of misinformation on their platforms by deleting misinformation \cite{BBC.2020}. While the Twitter API does not provide access to the content of source tweets that have been deleted, we were still able to analyze some of their characteristics. As part of an exploratory analysis, we found that \num{3663} potential rumor starter tweets have been deleted (either by Twitter or by the users themselves). An overwhelming majority of those (\SI{68.35}{\percent}) are potentially false rumors. Hence, even though these numbers suggest that a relevant proportion of false rumors on Twitter has been deleted, the vast majority of false rumors continue to circulate. For those rumors, our results demonstrate that falsehood can be more viral than the truth.

\textbf{Additional checks\footnote{Detailed results are reported in the Supplementary Materials.}: } Numerous exploratory analyses and checks validated our results and confirmed their robustness: (i) Since self-conscious emotions can be regarded as the counterpart of other-condemning emotions, we tested an alternative model specification in which we included the \emph{difference} between the emotion scores for other-condemning and self-conscious emotions instead of two individual variables. Consistent with our main analysis, we found that false rumors are more viral than the truth if the source tweet embeds a high proportion of other-condemning emotion words, whereas a high proportion of self-conscious emotion words is linked to a less viral spread. (ii) In our main analysis, we focused on rumors that are clearly true or false. However, \SI{17.7}{\percent}, of all rumors have been categorized as being of mixed veracity by the fact-checking organizations. We tested whether counting rumors of mixed veracity as either true or false affects the validity of our results. We find that our results are robust and that the combination of other-condemning emotion and mixed veracity is similarly viral as the combination of other-condemning emotions and false veracity. (iii) In our data, \SI{9.48}{\percent} of rumor starters have started more than one retweet cascade. To ensure that our models are not biased due to this source of non-independence, we dropped all users with clustering and reestimated the models. The results are robust and support our findings. We also repeated our analysis with monthly fixed effects to control for differences in the virality of rumor cascades due to different start dates. Also here, the results confirmed the findings from our main analysis. (iv) We repeated our analysis for subsets of rumor cascades that have been started by users that are either verified or not verified by Twitter. We find that our main findings hold for both user groups.

\section{Discussion}


Here we provide evidence that moral emotions play a crucial role in the spread of COVID-19 misinformation on social media. Using a comprehensive dataset of COVID-19 rumors that have been fact-checked by three independent fact-checking organizations (snopes.com, politifact.com, truthorfiction.com), we establish that other-condemning emotions -- also known as the hostility triad -- are linked to a more viral spread of false rumors.


While false rumors pose a threat to the successful overcoming of this pandemic, an understanding of how rumors diffuse in online social networks is -- even for non-crisis situations -- still in its infancy. Analyzing the spreading dynamics of fact-checked rumors is to a great extent generalizable to the spread of other (non-fact-checked) rumors on social media \cite{Vosoughi.2018}. Our finding that COVID-19 misinformation is, on average, more viral than the truth directly connects to the study from Vosoughi et al. \cite{Vosoughi.2018}, which yielded similar findings, yet outside the context of COVID-19. Previous research has also shown that misinformation on social media can have negative offline consequences. Among other instances, this has previously been confirmed to be the case during humanitarian crises \cite{Starbird.2014} and elections \cite{Aral.2019,Bakshy.2015,Grinberg.2019,Allcott.2017}. Our observation that COVID-19 misinformation is both widespread and viral on social media is at least equally concerning. COVID-19 misinformation not only poses severe health risks to individuals but also undermines the integrity of the political discourse \cite{Rapp.2018}. 


The results of this study highlight the role of moral emotions in rumor diffusion. Previous research \cite{Brady.2017} broadly distinguished moral vs.\ non-moral emotional expressions in social media content, while this work demonstrates that the two clusters of moral emotions (self-conscious vs.\ other-condemning emotions) have distinct effects on social transmission in the context of true and false rumors. We observe that false rumors receive more retweets than true rumors if the source tweets embed a high share of other-condemning emotions, whereas we find the opposite pattern, yet of smaller magnitude, for self-conscious emotions. Another relevant finding is that the expression of other-condemning emotion on virality is pronounced both for health misinformation and political misinformation. These findings may be partially explained by the high level of polarization of social media users in the context of COVID-19. In polarizing debates, radical ideas and beliefs are strengthened and more likely to translate into action. It thus seems plausible that the explosive mix of other-condemning emotions accelerates the spread of false rumors about those topics within social networks.


From a practical perspective, policy initiatives around the world urge social media platforms to limit the spread of false rumors \cite{Lazer.2018}. While previous research has studied emotions in {replies} to rumor cascades \cite{Prollochs.2021b,Prollochs.2021a}, our work highlights the importance of considering (moral) emotions in the {source tweets} that have initiated the rumor cascades. These findings could eventually be leveraged in machine learning models in order to detect false rumors more accurately. 
Emotion scores for source tweets are available immediately upon the beginning of the diffusion process -- a time point at which features from propagation dynamics are scarce \cite{Conti.2017}. Our findings may also be relevant with regard to other downstream tasks such as educational applications. Altogether, considering moral emotions in social media posts might help future works to develop more effective strategies against false rumors. 


This work is subject to the typical limitations of observational studies. We report associations and refrain from making causal claims. Future work should seek to corroborate our conclusions in controlled laboratory experiments and, in particular, test the causal influence of exposure to moral-emotional language on attitudes and behavior. Our inferences are also limited by the accuracy and availability of our data, specifically those from the three different fact-checking websites. For those, however, our data comprises all COVID-19 rumor cascades on Twitter until the end of April 2021. Despite these limitations, we believe that observing and understanding how misinformation spreads is the first step toward containing it. We hope that our work inspires more research into the causes, consequences, and potential countermeasures for the spread of misinformation -- both in crisis and non-crisis situations.

\section{Conclusion}

While false rumors pose a threat to the successful overcoming of this pandemic, an understanding of ``what makes false rumors viral'' is -- even for non-crisis situations -- still in its infancy. In this work, we approach this question through the lenses of morality and emotions and their role in rumor diffusion in polarized social media environments. For this purpose, we collected a unique dataset of COVID-19-related rumor cascades from Twitter and empirically analyze their spreading dynamics. We find that COVID-19 misinformation is, on average, more viral than the truth. However, the veracity effect is moderated by moral emotions: false rumors are more viral than the truth if the source tweets embed a high number of other-condemning emotion words, whereas a higher number of self-conscious emotion words is linked to a less viral spread. These findings offer insights into how true vs. false rumors spread and highlight the importance of considering moral emotions in social media content. 

\begin{acks}
This study was supported by a grant from the German Research Foundation (DFG grant 455368471). 
\end{acks}


\bibliographystyle{ACM-Reference-Format-no-doi-abbrv}
\balance
\bibliography{literature}

\clearpage
\appendix

\begin{center}
\LARGE
\textbf{Supplementary Materials}
\end{center}
\normalsize

\section{Topic modeling}

Table~\ref{tbl:seed_words} shows the manually selected seed words that were used to identify topic-related tweets in weakly supervised learning. 

Table~\ref{tbl:seed_words} reports the results for our user study testing for the presence of errant tweets. On average, the share of tweets that were not classified as at least ``somewhat related to  [topic]'' was lower than {\SI{8.5}{\percent}}. 

\begin{table}
\footnotesize
	\begin{center}
		\caption{Seed words used to identify topic-related tweets in weakly supervised learning. Various word forms of the keywords are also considered, \eg, \enquote{masks} and \enquote{masking} are also considered for the keyword \enquote{mask}.}
		\begin{tabular}{lp{.8\columnwidth}}
			\toprule
			{Topic}           & {Seed keywords}                                                                                 \\
			\midrule
			\textsc{Politics} & Bill, Trump, Biden, Obama, Democrats, GOP, Republicans, Tax, Administration, Red, Blue, Pelosi, Economy, Chinavirus \\
			\textsc{Health}   & Vaccine, Flu, Mask, Fever, Ebola, SARS, Ibuprofen, Garlic, Health, Infection                    \\ \bottomrule
		\end{tabular}
		\label{tbl:seed_words}
	\end{center}
\end{table}

\begin{table}
\footnotesize
	\begin{center}
		\caption{Frequency of errors in topic labeling.}
		\begin{tabular}{lr}
			\toprule
			\multicolumn{1}{l}{{Topic}} & \multicolumn{1}{r}{{Percent Error}} \\
			\midrule
			\textsc{Politics}           & 6.0\%                               \\
			\textsc{Health}             & 2.2\%                               \\
			\textsc{Other}              & 17.5\%                                \\ \midrule
			Mean & 8.5\% \\
			\bottomrule
		\end{tabular}
		\label{tbl:labeling}
	\end{center}
\end{table}

\section{Analysis of control variables}

We tested a model specification in which we only incorporated control variables from previous works. Table~\ref{tbl:models_noint} shows that rumors receive a particularly high number of retweets if they are false and if they have been started by users with a larger number of followers.

\begin{table}[H]
\footnotesize
	\begin{center}
		\caption{Regression results for control variables only. The dependent variable is the number of retweets.}
		\begin{tabularx}{\columnwidth}{@{\hspace{\tabcolsep}\extracolsep{\fill}}l S}
			\toprule
			\multicolumn{2}{l}{Dependent Variable: $\mathit{RetweetCount}$}                                                                                                                                               \\
			\midrule
			Falsehood                     & 0.145^{**}                         \\
			                              & (0.050)                            \\
			Followers                     & 1.134^{***}                        \\
			                              & (0.036)                            \\
			Followees                     & 0.046                              \\
			                              & (0.025)                            \\
			AccountAge                    & -0.217^{***}                       \\
			                              & (0.026)                            \\
			HasMedia                      & -0.132^{*}                         \\
			                              & (0.052)                            \\
			Verified                      & 0.007                              \\
			                              & (0.073)                            \\
			Intercept                     & 7.250^{***}                        \\
			                              & (0.058)                            \\
			\midrule
			Observations (rumor cascades) & {\num{8727}}                       \\
			\bottomrule
			\multicolumn{2}{l}{Sign. levels: $^*p< 0.05$, $^{**}p< 0.01$, $^{***}p< 0.001$; standard errors in parentheses}                                                                                        \\
		\end{tabularx}
		\label{tbl:models_noint}
	\end{center}
\end{table}

\section{Verified vs. unverified users}

We repeated our analysis for subsets of rumor cascades that have been started by users that are verified or not-verified by Twitter. Table~\ref{tbl:models_ver_unver} shows that our main findings hold for both user groups.

\begin{table}[H]
\footnotesize
	\begin{center}
		\caption{Regression results for rumor cascades initiated from \textsc{Verified} (column 1) or \textsc{Non-verified} (column 2) users only.}
		\begin{tabularx}{\columnwidth}{@{\hspace{\tabcolsep}\extracolsep{\fill}}l *{2}{S} }
			\toprule
			\multicolumn{3}{l}{Dependent Variable: $\mathit{RetweetCount}$}                                                                        \\
			\midrule
			                                   & \multicolumn{1}{c}{Subset: \textsc{Verified}} & \multicolumn{1}{c}{Subset: \textsc{Non-verified}} \\
			\midrule
			Falsehood $\times$ OtherCondemning & 0.238^{***}                                   & 0.302^{**}                                        \\
			                                   & (0.058)                                       & (0.100)                                           \\
			Falsehood $\times$ SelfConscious   & -0.256^{***}                                  & -0.262^{*}                                        \\
			                                   & (0.055)                                       & (0.111)                                           \\
			Falsehood                          & 0.182^{***}                                   & 0.062                                             \\
			                                   & (0.049)                                       & (0.104)                                           \\
			OtherCondemning                    & -0.016                                        & -0.025                                            \\
			                                   & (0.042)                                       & (0.085)                                           \\
			SelfConscious                      & 0.100^{**}                                    & 0.073                                             \\
			                                   & (0.037)                                       & (0.096)                                           \\
			Followers                          & 0.871^{***}                                   & 2.138^{***}                                       \\
			                                   & (0.043)                                       & (0.075)                                           \\
			Followees                          & 0.101^{***}                                   & -0.512^{***}                                      \\
			                                   & (0.027)                                       & (0.055)                                           \\
			AccountAge                         & -0.090^{*}                                    & -0.265^{***}                                      \\
			                                   & (0.036)                                       & (0.041)                                           \\
			HasMedia                           & -0.344^{***}                                  & 0.142                                             \\
			                                   & (0.053)                                       & (0.104)                                           \\
			Intercept                          & 7.412^{***}                                   & 7.969^{***}                                       \\
			                                   & (0.048)                                       & (0.106)                                           \\
			\midrule
			Observations (rumor cascades)      & {\num{4836}}                                  & {\num{3891}}                                      \\
			\bottomrule
			\multicolumn{3}{l}{Sign. levels: $^*p< 0.05$, $^{**}p< 0.01$, $^{***}p< 0.001$; standard errors in parentheses}                 \\
		\end{tabularx}
		\label{tbl:models_ver_unver}
	\end{center}
\end{table}

\section{Rumors with mixed veracity}

\begin{table}[H]
\footnotesize
	\begin{center}
		\caption{Regression results with mixed rumors categorized as false rumors (Model 1) and mixed rumors categorized as true rumors (Model 2).}
		\begin{tabularx}{\columnwidth}{@{\hspace{\tabcolsep}\extracolsep{\fill}}l *{2}{S} }
			\toprule
			\multicolumn{3}{l}{Dependent Variable: $\mathit{RetweetCount}$}                                                        \\
			\midrule
			                                   & \multicolumn{1}{c}{Model (1)} & \multicolumn{1}{c}{Model (2)}                     \\
			\midrule
			Falsehood $\times$ OtherCondemning & 0.298^{***}                   & 0.165^{***}                                       \\
			                                   & (0.046)                       & (0.050)                                           \\
			Falsehood $\times$ SelfConscious   & -0.341^{***}                  & -0.100^{*}                                        \\
			                                   & (0.046)                       & (0.049)                                           \\
			Falsehood                          & 0.085                         & 0.156^{***}                                       \\
			                                   & (0.044)                       & (0.046)                                           \\
			OtherCondemning                    & -0.054                        & -0.026                                            \\
			                                   & (0.033)                       & (0.042)                                           \\
			SelfConscious                      & 0.163^{***}                   & 0.076                                             \\
			                                   & (0.033)                       & (0.040)                                           \\
			Followers                          & 1.175^{***}                   & 1.173^{***}                                       \\
			                                   & (0.033)                       & (0.033)                                           \\
			Followees                          & 0.039                         & 0.031                                             \\
			                                   & (0.022)                       & (0.022)                                           \\
			AccountAge                         & -0.125^{***}                  & -0.132^{***}                                      \\
			                                   & (0.024)                       & (0.024)                                           \\
			HasMedia                           & -0.122^{*}                    & -0.131^{**}                                       \\
			                                   & (0.048)                       & (0.048)                                           \\
			Verified                           & -0.085                        & -0.077                                            \\
			                                   & (0.066)                       & (0.066)                                           \\
			Intercept                          & 7.344^{***}                   & 7.278^{***}                                       \\
			                                   & (0.049)                       & (0.054)                                           \\
			\midrule
			Observations (rumor cascades)      & {\num{10610}}                 & {\num{10610}}                                     \\
			\bottomrule
			\multicolumn{3}{l}{Sign. levels: $^*p< 0.05$, $^{**}p< 0.01$, $^{***}p< 0.001$; standard errors in parentheses} \\
		\end{tabularx}
		\label{tbl:models_fulver_mixfal}
	\end{center}
\end{table}

In our main analysis, we focused on rumors that are clearly true or false. However, \SI{17.7}{\percent}, of all rumors have been categorized as being of mixed veracity by the fact-checking organizations. We tested whether counting rumors of mixed veracity as either true or false affects the validity of our results. Table~\ref{tbl:models_fulver_mixfal} shows that our results are robust and that the combination of other-condemning emotion and mixed veracity is similarly viral as the combination of other-condemning emotions and false veracity.


\section{Sensitivity to non-independence}

In our data, \SI{9.48}{\percent} of rumor starters have started more than one retweet cascade. To ensure that our models are not biased due to this source of non-independence, we dropped all users with clustering and reestimated the models. Table~\ref{tbl:models_mfe_crse} show that the results are robust and support our findings.

We also repeated our analysis with monthly fixed effects to control for differences in the virality of rumor cascades due to different start dates (Table~\ref{tbl:models_mfe_crse}). All results confirm the findings from our main analysis.

\begin{table}[H]
\footnotesize
	\begin{center}
		\caption{Regression results without rumor cascades from users that have started more than one retweet cascade (Model 1) and with monthly fixed effects (Model 2).}
		\begin{tabularx}{\columnwidth}{@{\hspace{\tabcolsep}\extracolsep{\fill}}l *{2}{S} }
			\toprule
			\multicolumn{3}{l}{Dependent Variable: $\mathit{RetweetCount}$}                                                        \\
			\midrule
			                                   & \multicolumn{1}{c}{Model (1)} & \multicolumn{1}{c}{Model (2)}                     \\
			\midrule
			Falsehood $\times$ OtherCondemning & 0.428^{***}                   & 0.225^{***}                                       \\
			                                   & (0.100)                       & (0.052)                                           \\
			Falsehood $\times$ SelfConscious   & -0.386^{***}                  & -0.245^{***}                                      \\
			                                   & (0.104)                       & (0.052)                                           \\
			Falsehood                          & 0.118                         & 0.226^{***}                                       \\
			                                   & (0.101)                       & (0.051)                                           \\
			OtherCondemning                    & -0.101                        & 0.005                                             \\
			                                   & (0.085)                       & (0.041)                                           \\
			SelfConscious                      & 0.128                         & 0.109^{**}                                        \\
			                                   & (0.089)                       & (0.040)                                           \\
			Followers                          & 1.937^{***}                   & 1.289^{***}                                       \\
			                                   & (0.081)                       & (0.036)                                           \\
			Followees                          & -0.235^{***}                  & 0.061^{*}                                         \\
			                                   & (0.058)                       & (0.024)                                           \\
			AccountAge                         & -0.246^{***}                  & -0.273^{***}                                      \\
			                                   & (0.043)                       & (0.026)                                           \\
			HasMedia                           & 0.254^{*}                     & -0.162^{**}                                       \\
			                                   & (0.099)                       & (0.052)                                           \\
			Verified                           & -0.326^{*}                    & 0.034                                             \\
			                                   & (0.137)                       & (0.072)                                           \\
			Intercept                          & 7.857^{***}                   & 7.445^{***}                                       \\
			                                   & (0.122)                       & (0.586)                                           \\
			\midrule
			Monthly fixed effects              & {\xmark}                        & {\checkmark}                                        \\
			\midrule
			Observations (rumor cascades)      & {\num{4139}}                  & {\num{8727}}                                      \\
			\bottomrule
			\multicolumn{3}{l}{Sign. levels: $^*p< 0.05$, $^{**}p< 0.01$, $^{***}p< 0.001$; standard errors in parentheses} \\
		\end{tabularx}
		\label{tbl:models_mfe_crse}
	\end{center}
\end{table}

%
\section{Alternative emotion measure}

Since self-conscious emotions can be regarded as the counterpart of other-condemning emotions, we tested an alternative model specification in which we included the \emph{difference} between the emotion scores for other-condemning and self-conscious emotions instead of two individual variables (Table~\ref{tbl:models_diff}). Consistent with our main analysis, we found that false rumors are more viral than the truth if the source tweet embeds a high proportion of other-condemning emotion words, whereas a high proportion of self-conscious emotion words is linked to a less viral spread.

\begin{table}[H]
\footnotesize
	\begin{center}
		\caption{Retweet count as a function of the difference of other-condemning and self-conscious emotions (OtherCondemning--SelfConscious).}
		\begin{tabularx}{\columnwidth}{@{\hspace{\tabcolsep}\extracolsep{\fill}}l S}
			\toprule
			\multicolumn{2}{l}{Dependent Variable: $\mathit{RetweetCount}$}                                                                                                                                                      \\
			\midrule
			Falsehood $\times$                   & 0.289^{***}                        \\
			\quad OtherCondemning--SelfConscious & (0.048)                            \\
			Falsehood                            & 0.127^{*}                          \\
			                                     & (0.050)                            \\
			OtherCondemning--SelfConscious       & -0.051                             \\
			                                     & (0.038)                            \\
			Followers                            & 1.161^{***}                        \\
			                                     & (0.036)                            \\
			Followees                            & 0.057^{*}                          \\
			                                     & (0.025)                            \\
			AccountAge                           & -0.196^{***}                       \\
			                                     & (0.026)                            \\
			HasMedia                             & -0.107^{*}                         \\
			                                     & (0.052)                            \\
			Verified                             & -0.035                             \\
			                                     & (0.073)                            \\
			Intercept                            & 7.260^{***}                        \\
			                                     & (0.058)                            \\
			\midrule
			Observations (rumor cascades)        & {\num{8727}}                       \\
			\bottomrule
			\multicolumn{2}{l}{Sign. levels: $^*p< 0.05$, $^{**}p< 0.01$, $^{***}p< 0.001$; standard errors in parentheses}                                                                                               \\
		\end{tabularx}
		\label{tbl:models_diff}
	\end{center}
\end{table}

\end{document}